\documentclass[12pt,runningheads]{llncs}
\usepackage[utf8]{inputenc}
\usepackage{fullpage}
\usepackage{changepage}
\usepackage{color}
\usepackage{csquotes}
\usepackage{amsmath}
\usepackage{graphicx}
\usepackage{xkeyval}
\usepackage{xstring}
\usepackage{iftex}
\usepackage{hyperref}
\newcommand{\CSD}{CSD}

\title{\textbf{More style, less work: card-style data decrease risk-limiting audit sample sizes}}

\author{Amanda K. Glazer, Jacob V. Spertus, and Philip B. Stark}
\institute{
University of California, Berkeley, Department of Statistics\\
\email{amandaglazer@berkeley.edu}; \email{jakespertus@berkeley.edu};
\email{pbstark@berkeley.edu}
}
\date{December 2020}

\begin{document}

\maketitle

\begin{abstract}
    U.S.\ elections rely heavily on computers such as voter registration databases, electronic pollbooks, voting machines, scanners, tabulators, and results reporting websites.
    These introduce digital threats to election outcomes. 
    Risk-limiting audits (RLAs) mitigate threats to some of these systems
    by manually inspecting random samples of ballot cards.
    RLAs have a large chance of correcting wrong outcomes (by conducting a full manual tabulation of a trustworthy record of the votes), 
    but can save labor when reported outcomes are correct. 
    This efficiency is eroded when sampling cannot be targeted to ballot cards that contain the contest(s) under audit. 
    If the sample is drawn from all cast cards, RLA sample sizes scale like the reciprocal of the fraction of ballot cards that contain the contest(s) under audit.
    That fraction shrinks as the number of cards per ballot grows (i.e.,
    when elections contain more contests) and as the fraction of ballots that contain the contest decreases (i.e., when a smaller percentage of voters
    are eligible to vote in the contest).
    States that conduct RLAs of contests on multi-card ballots or of small contests can dramatically reduce sample sizes by using information
    about which ballot cards contain which contests---by
    keeping track of card-style data (\CSD). 
    For instance, \CSD\ reduces the expected number of draws needed to audit a single countywide contest on a 4-card ballot by 75\%. 
    Similarly, \CSD\ reduces the expected number of draws by 95\% or more
    for an audit of two contests with the same margin on a 4-card ballot if one contest is on every ballot and the other is on 10\% 
    of ballots.
    In realistic examples, the savings can be several
    orders of magnitude.
\end{abstract}

\begin{adjustwidth}{2cm}{2cm}
Style is a way to say who you are without having to speak.

\hfill -- Rachel Zoe

\end{adjustwidth}

\section{Introduction}

The principle of \emph{evidence-based elections} is that elections should provide convincing evidence that the reported winners really won \cite{starkWagner12}. 
Evidence-based elections require a trustworthy record of the votes.
Generally, that means hand-marked paper ballots (with an accessible marking option for voters who require accommodations) kept demonstrably secure throughout the canvass \cite{appelStark20,appelEtal20}.
Crucially, the reported results must be checked against that trustworthy paper trail. 
\textit{Risk-limiting audits} (RLAs) provide a rigorous way to perform that check so that there is a large chance of correcting the reported outcome if it is wrong.
The probability that an RLA does not correct the reported outcome if the reported outcome is wrong is less than the risk limit 
\cite{stark08a,lindemanStark12}. 

U.S.~elections generally include several to dozens of contests that widely vary in size: each voter is eligible to vote in a subset of those contests.
For example, a voter may be eligible to vote for president, senators, governor, representatives, mayor, judges, school board seats, local tax measures, etc. 
Because ballots often contain so many contests, they 
generally comprise more than one \emph{card} or \emph{sheet} or \emph{page},
each containing some of the contests.

The term \emph{ballot style} generally refers to the set of contests on a given voter's ballot. 
(Ballot style can also encode precinct information, i.e.,
even if voters in two different precincts are eligible to vote in the same set of contests, ballots for the two precincts are considered to be of two different styles.) 
Here, we use \emph{card style}
to refer to the set of contests on a given ballot card, and \CSD\ to refer to
card-style data for an election.

RLAs generally involve inspecting a random sample of ballot cards. 
States that perform RLAs have so far drawn the sample of ballot cards to inspect from all the cards cast in the election, without regard for the contests those cards contain.
When a contest under audit appears on only a small fraction of ballot cards, this can make the audit unnecessarily costly. 
Especially when the
margin\footnote{%
Technically, the \emph{diluted margin} 
\cite{stark10d}
drives sample sizes for ballot-level comparison audits,
as described below.
The diluted margin is the margin in votes divided by the total number of cards in the population from which the sample is drawn.
It is generally less than the conventional margin, which is the margin in votes divided by the number of valid votes in the contest,
excluding undervotes and overvotes.
}
is small, an RLA based on a sample drawn from all ballot
cards requires manually inspecting far more cards than an RLA that only samples cards that contain the
contest.
Targeting the sample to just those cards is possible if
card-style data are available, that is, a listing of the contests each cast card contains.

A similar issue arises for large contests---even jurisdiction-wide contests---when ballots consist of multiple cards: if the sample can be drawn just from cards that contain those contests, a much smaller sample
may suffice to confirm the outcome
than if the sample is drawn indiscriminately from all cast cards
\cite{lindemanEtal18}.
As a rule of thumb, if a contest is on a fraction $f$ of ballot cards cast in the contest, the sample size required to confirm the contest outcome will be roughly $1/f$ times larger than if the sample can be drawn just from cards that contain the contest.

Here, we show that \CSD\---i.e., keeping track of which cards contain which contests---can reduce audit sample sizes by orders of magnitude in typical elections. 
Section~\ref{sec:background} provides additional context and defines key terms.
Section~\ref{sec:1cardballots} examines the simplest case with a ballot-level comparison audit: auditing two contests in an election with a one-card ballot.
One contest is on every card; the other is on only some of the cards.
Section~\ref{sec:multicardballots} considers auditing two contests in an election with multi-card ballots. 
Section~\ref{sec:bpa} extends our analysis to 
ballot-polling audits. 
Section~\ref{sec:case_studies} presents case studies of hypothetical audits in two California counties of different sizes.
Section~\ref{sec:implementation} sketches how to implement an audit that takes advantage of \CSD\ 
using \emph{consistent sampling}.
Section~\ref{sec:conclusions} presents conclusions and recommendations.


\section{Background}
\label{sec:background}
\subsection{Ballots, cards, ballot manifests, and card styles}
\label{sec:ballots_and_cards}
A \textit{ballot} is what the voter receives and casts; a \textit{ballot card} is an individual page of a ballot. 
In the U.S., ballots often consist of more than one card.
The ballot cards that together comprise a ballot generally do not stay together once they are cast.
RLAs generally draw ballot \emph{cards} at random---not ``whole'' ballots.

To conduct an RLA, an upper bound on the 
number of 
validly cast 
ballot cards must be known before the audit begins.
The bound could come from manually keeping track of the paper, or from other information available to the election official, such as the number of voters eligible to vote in each contest, the number of pollbook signatures, or the number of ballots sent to polling places, mailed to voters, and returned by voters \cite{banuelosStark12}.

RLAs generally rely on \emph{ballot manifests} to draw
a random sample of ballot cards.
A ballot manifest describes how the physical ballot cards are stored.
It is the \emph{sampling frame} for the audit.
This paper explains how it can be beneficial to augment
the ballot manifest with information about the \emph{style}
of each card, i.e., the particular contests the card
contains---card-style data (\CSD).
Until recently,\footnote{In November 2019, a pilot RLA was conducted in
San Francisco that used \CSD\ \cite{blomEtal20}.
}
\CSD\ has not been used in RLAs.
Figures~\ref{fig:1page_CSD} and \ref{fig:multicard_CSD} respectively display examples of \CSD\ for single-card and multi-card ballots.

\begin{figure}[!ht]
    \centering
     \begin{tabular}{|l|l|l|l|l|l|}
        \hline
        \textbf{Cart} & \textbf{Tray} & \textbf{Position in Tray} & \textbf{Governor} & \textbf{Mayor of Irvine} & ...\\ 
        \hline
        1 & 4 & 96 & Yes & No & ... \\
        \hline
        5 & 1 & 12 & Yes & No & ... \\
        \hline 
        2 & 2 & 72 & Yes & No & ... \\
        \hline
        ... & ... & ... & ... & ... & ... \\
        \hline
        3 & 5 & 50 & Yes & Yes & ...\\
        \hline
    \end{tabular}
    \caption{A hypothetical example of card-style data (\CSD) for an election in Orange County California with one-card ballots. Ballot cards are uniquely identified by their position (cart, tray, position in tray). \CSD\ further identifies contests that each ballot card is supposed to contain (truncated to two contests here), appearing as columns. Here we display the records for the county-wide Governor's race and the race for Mayor of Irvine (a city within Orange County). More storage-efficient \CSD\ might associate an unstructured list to each ballot card with numeric identifiers of the contests it contains (e.g., $\{1,4,10,12\}$). The public should be able to check that there are $N$ lines in the \CSD, where $N$ is the number of ballots (and ballot cards) cast in the county. If $N_S$ ballots were cast in contest $S$, there should be $N_S$ lines with ``yes'' in the column corresponding to contest $S$ in the \CSD.}
    \label{fig:1page_CSD}
\end{figure}

\begin{figure}[!ht]
    \centering
    \begin{tabular}{|l|l|l|l|l|l|}
        \hline
        \textbf{Cart} & \textbf{Tray} & \textbf{Position in Tray} & \textbf{Governor} & \textbf{Mayor of Irvine} & ...\\ 
        \hline
        2 & 6 & 3 & No & Yes & ... \\
        \hline
        5 & 1 & 12 & No & No & ... \\
        \hline 
        2 & 5 & 64 & Yes & No & ... \\
        \hline
        ... & ... & ... & ... & ... & ... \\
        \hline
        1 & 2 & 8 & Yes & No & ...\\
        \hline
    \end{tabular}
    \caption{An example of card-style data (\CSD) for a hypothetical election in Orange County with multi-card ballots. The Governor's race and the race for Mayor of Irvine appear on different cards. Because neither contest can appear on a card in this example (e.g., line 2), there must be at least 3 card styles.}
    \label{fig:multicard_CSD}
\end{figure}

There are two principal ways to generate \CSD: (1)~physically sort the ballot cards according to the contests they contain, or (2)~rely on the voting 
system for that information---even though it
might not be accurate.
Precinct-based voting partially sorts ballot cards: If every voter in the precinct is eligible to vote in the same contest(s) and the ballot has only one card, then each precinct's ballot cards have a single style.
Knowing which precinct a bundle of ballot cards came from
then tells us the contests on each card.
This does not work for multi-card ballots.
Some jurisdictions sort vote-by-mail ballots by precinct before scanning, which also partially sorts the cards.
Vote centers make sorting ballots more difficult because each center receives ballot styles from more than one precinct, and ballots cast in vote centers generally are not sorted before they are scanned.

Some modern vote tabulation systems record a cast-vote record (CVR, a record of how the voting system interpreted the selections on the ballot card)
for each ballot card in a way that allows the corresponding physical card to be identified and retrieved, and vice versa.
Such systems are amenable to efficient RLAs and they 
also contain (possibly inaccurate) \CSD:
the contests on a card can be inferred from its CVR, as long as the CVR encodes ``no selection'' for contests in which the voter did not express a preference, according to the voting system.
\CSD\ derived from CVRs rely on the voting system, so they could be wrong: \CSD\ might show that a card contains a contest it does not contain, or vice versa.
Such errors can be accounted for rigorously in the audit using the ``manifest phantoms to evil zombies'' approach \cite{banuelosStark12,stark20a}, described below.
The same approach can accommodate errors
in \CSD\ uncovered in auditing
manually or machine-sorted ballot cards.

Because physically sorting cards is expensive,
we expect that \CSD\ generally will 
not be available unless
the jurisdiction has a voting system that can
produce CVRs linked to physical ballots.
If such CVRs are 
available, 
\emph{ballot-level comparison} RLAs are possible.
Such RLAs are especially efficient, so we emphasize them below. 

\subsection{Ballot-polling and ballot-level comparison audits}
\label{sec:polling_and_comparison_audits}

There are two common approaches to auditing: \emph{comparison}
and \emph{ballot-polling} \cite{lindemanStark12}
(there are also audits that combine the two approaches;
see, e.g., \cite{ottoboniEtal18,stark20a}).
Comparison audits involve comparing manual tabulation of physically identifiable sets of ballot cards to the machine tabulation of the same ballot cards.
The efficiency of comparison audits increases as the size of the sets decreases.
The most efficient comparison audits compare human interpretation of individual ballot cards to the machine interpretation of individual ballot cards, CVRs. 
Such audits are called \emph{ballot-level comparison audits}
(in contrast to \emph{batch-level comparison audits}, which compare the manual and electronic tabulation of batches of ballots, such as ballots cast in person in a particular precinct).
Ballot-level comparison audits are possible only if the voting system produces CVRs
that can be linked to the corresponding physical ballot card.\footnote{%
There are also \emph{transitive} ballot-level comparison audits,
which involve re-scanning the
ballot cards using an unofficial system.
}
Legacy voting systems generally do not, but many
newer systems do. 

Ballot-polling audits check the outcome
but do not check or rely on the machine tabulation.
They do not require CVRs or machine subtotals: all they require
is paper ballots, organized well enough to sample cards at random.

Here, we focus on ballot-level comparison audits but we briefly address ballot-polling audits.
We do not address batch-level comparison audits.

Ballot-level comparison audits and ballot-polling audits sample individual physical ballot cards.
We refer to the act of sampling a single ballot card as a \textit{draw}. 
Retrieving and inspecting ballot cards is labor intensive, so when the reported outcome is correct we want to minimize the number of draws (i.e., the sample size).
We show below that knowing which cards contain which
contests can dramatically reduce the number of draws
required to confirm correct outcomes.
(When the outcome is incorrect, we \emph{want} the audit
to inspect every ballot, in order to determine
the correct outcome.)

\subsection{Super-simple simultaneous single-ballot RLAs}
\label{sec:s4}
In this paper we use the super-simple simultaneous single-ballot (S4) RLA of \cite{stark10d} to 
illustrate workloads. 
Although S4 is not the most efficient RLA method, 
in part because it relies on sampling with replacement, it allows simple workload computations without the need for simulation.
Moreover, if the sample size is small relative to the 
number of ballots cast and the number of observed discrepancies is small, the method is close to the best known.
(If the sample size is an appreciable fraction of the population, other methods can be far more efficient.
See, e.g., \cite{stark20a}.)
We expect that the savings in workload afforded by \CSD\ 
will be substantial for all RLA methods.

The number of draws S4 needs to confirm results depends on the diluted margin and
the number and nature of discrepancies the sample uncovers.\footnote{%
The S4 method has one tuning parameter, $\gamma$, which does not affect the risk limit but does affect the workload.
To estimate the final workload, we assume
that some fraction of the ballots in the
sample will reveal 1-vote
overstatements (errors that inflated a reported winner's margin over a reported loser by 1 vote).
See \cite{stark10d}.
}
The initial sample size can be written as a constant (denoted $\rho$)\footnote{%
In general, $\rho = -\log(\alpha)/[\frac{1}{2\gamma} + \lambda \log (1 - \frac{1}{2\gamma})]$, where $\gamma$ is an error inflation factor and $\lambda$ is
the anticipated rate of one-vote
overstatements in the initial sample as a percentage
of the diluted margin \cite{stark10d}. 
We define $\gamma$ and $\lambda$ as in \url{https://www.stat.berkeley.edu/~stark/Vote/auditTools.htm}.} divided by the ``diluted margin.'' 
With \CSD{}, there are two relevant ``diluted margins,'' as we shall see. 
The \emph{partially diluted margin}
is the margin in votes divided by the number of \emph{cards} that contain the contest, including cards with undervotes or no valid vote in the contest.
This differs from how margins are often reported, where the denominator is only the \emph{valid votes} in the contest, not the number of cards cast in the contest.
The \emph{fully diluted margin} is the margin in votes divided by the number of cards in the population of cards from which the audit sample is drawn.
When the sample is drawn only from cards that contain the contest, the partially diluted margin and the fully diluted margin are equal; otherwise, the fully diluted margin is smaller.
If \CSD\ are unavailable, the number of cards in that population is the number 
of cards cast in the jurisdiction.
If \CSD\ are available, the number of cards in the population can be reduced to the number of cards that contain the contest. 
The availability of \CSD\ drives the sample size through the difference
between the partially and fully diluted margins. 

\section{One-Card Ballots}
\label{sec:1cardballots}

Consider an election with a one-card ballot.
We want to audit two contests:
a jurisdiction-wide contest $B$ (for ``big'') listed
on every card and a smaller contest $S$ (for ``small'') 
listed on some of the cards. 
For example, $B$ might be a countywide contest such as sheriff and $S$ might be a mayoral race. 

There are $N$ ballots cast in the jurisdiction, of which $N_B = N$ contain contest $B$ and $N_S = pN < N$ contain contest $S$, where 
$p \in (0, 1)$.\footnote{The fraction $p$ can be quite small in real elections. 
For instance, a 2018 ballot measure in San Mateo County, California, had 687~eligible voters out of 399,591 registered voters,
i.e., $p \approx 0.0017$.}

The reported margin of contest $B$ is $M_B$ votes and the reported margin of contest $S$ is $M_S$ votes.
Let $m_B \equiv M_b/N_B$ and $m_S \equiv M_S/N_S$
be the two partially diluted margins.
We assume that the ballot manifest and the CVR both indicate that $N$ ballots were cast overall, and that ballot cards are where the manifest says they are: there are no ``phantoms'' in the terminology of \cite{banuelosStark12}. 
We find sample sizes for a risk limit of $0.05$
on the assumption that the rate of one-vote overstatements will be 0.001.
We assume that other types of errors did not occur. 

Absent \CSD, the sample for auditing contest $S$ would be drawn from the entire population of $N$ ballots.
Contest $S$ is on $pN$ cards, so, in addition to the cards that contain undervotes or invalid votes in contest $S$, there are in effect another $(1-p)N$ cards with non-votes in contest $S$. 
The fully diluted margin for contest $S$ 
is thus $M_S/N = p m_S$. 
For simplicity, we use the same risk limit for both contests (5\% in our numerical examples).
The number of cards we need to sample will be the smaller of the sample size for contest $B$ (with fully diluted margin $m_B$ and risk limit $\alpha$), and for contest $S$ (with fully diluted margin $p m_S$ and risk limit $\alpha$). 
If $m_B \leq p m_S$ then we will be done auditing both contests when the audit of contest $B$ is complete (assuming the number of discrepancies was not larger than anticipated). 
However, if $m_B > p m_S$ we must sample more ballots to finish the audit of contest $S$.

Here's an example.
Suppose $N = \mbox{10,000}$, 
$p = 0.1$, and $m_B = 0.1 = m_S$. 
For contest $B$, there are 5,500 reported votes for the winner and 4,500 reported votes for the loser; for contest $S$ there are 550 votes for the winner and 450 votes for the loser. 
Contest $B$ has an initial sample size of 64~cards, and contest $S$ has an initial sample size of 721~cards.
Therefore, we will need to sample 721 cards. 
In general, for contest $S$, we would need to sample approximately $1/p$ times more cards than if we sampled just from the cards that contain contest $S$. 
When $m_S$ and $p$ are small, this can lead to very large samples. 
For example, if $p = 0.01$ and $n$ is the number of cards we would need to inspect if we were sampling just from cards that contained the contest with partially diluted margin $m_S$, 
then we could need to sample approximately $100n$ cards if we sampled from all cards (not just those that contain contest $S$).

If there are \CSD, the sample can be substantially
smaller:
We could first sample from all the cards until the audit of contest $B$ is complete. 
Then we can use \CSD\ to draw additional cards that contain contest $S$.

Consider the same example as before with $N = \mbox{10,000}$, $p = 0.1$, and $m_B = 0.1 = m_S$, but suppose we have \CSD. 
To audit $B$ we still need to draw 64~cards.
We expect that $64p = 6.4 \approx 6$ of those cards 
will also contain contest $S$ and $64(1-p) \approx 58$ will not. 
To finish the audit of $S$, we expect to need to draw 58 more cards, all containing contest $S$, for a
total of 122~cards. 
\CSD\ reduces the number of audited
cards from 721 to about 122, a factor of almost~6.

\begin{figure}[!ht]
    \centering
    \includegraphics[width = \textwidth]{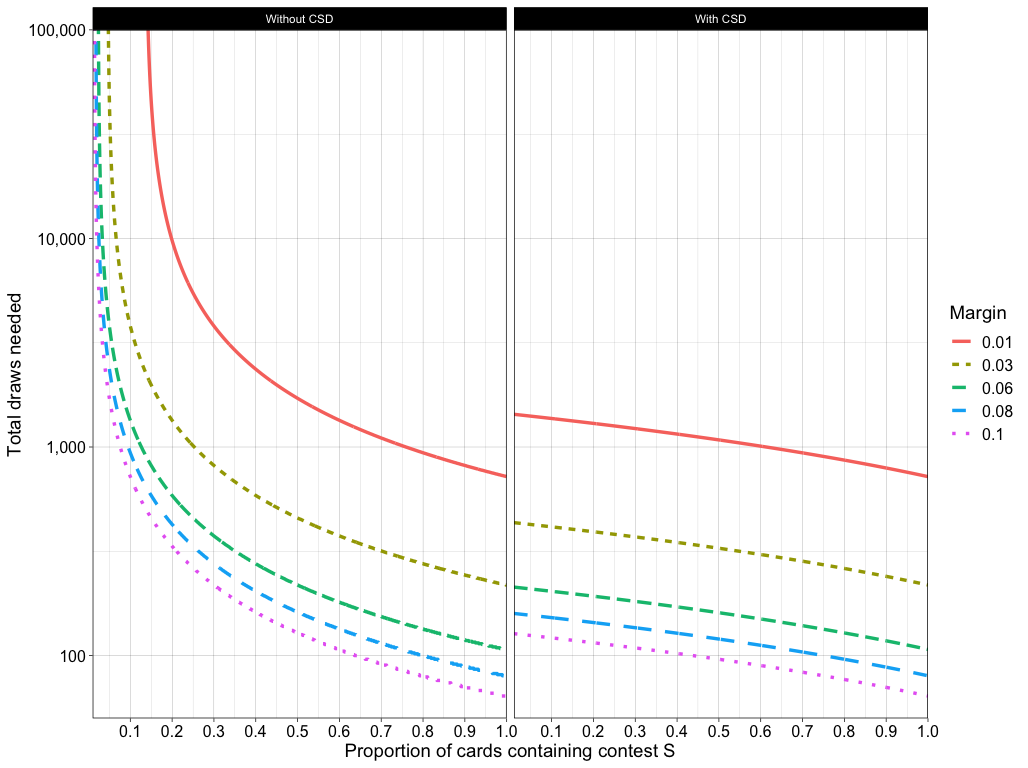}
    \caption{Expected draws needed to confirm the outcomes in both contest $B$ and contest $S$ when the reported outcomes are correct with partially diluted margin $m_B = m_S$ given in the legend. An error rate of 0.001 1-vote overstatements is assumed. 
    The $y$-axis is the expected number of draws (i.e., the sample size) needed on the $\log_{10}$ scale and is truncated at 100,000 draws. 
    The left panel gives the expected number of draws if \CSD\ are unavailable, while the right panel gives the expected number of draws with \CSD. The $x$-axis ranges over a grid of proportions $p$ of ballot cards containing the small contest, from 1 in 100 ($p = 0.01$) to every ballot card ($p = 1.0$). When contest $S$ is on every card ($p = 1$), the workload is the same with or without \CSD.}
    \label{fig:equal_margin}
\end{figure}

Figure~\ref{fig:equal_margin} displays total ballot cards needed to audit contests $B$ and $S$, with and without \CSD, across a range of partially diluted margins $m_B = m_S$ and proportions $p$ of cards on which contest $S$ appears.

\begin{figure}[!ht]
    \centering
    \includegraphics[width = \textwidth]{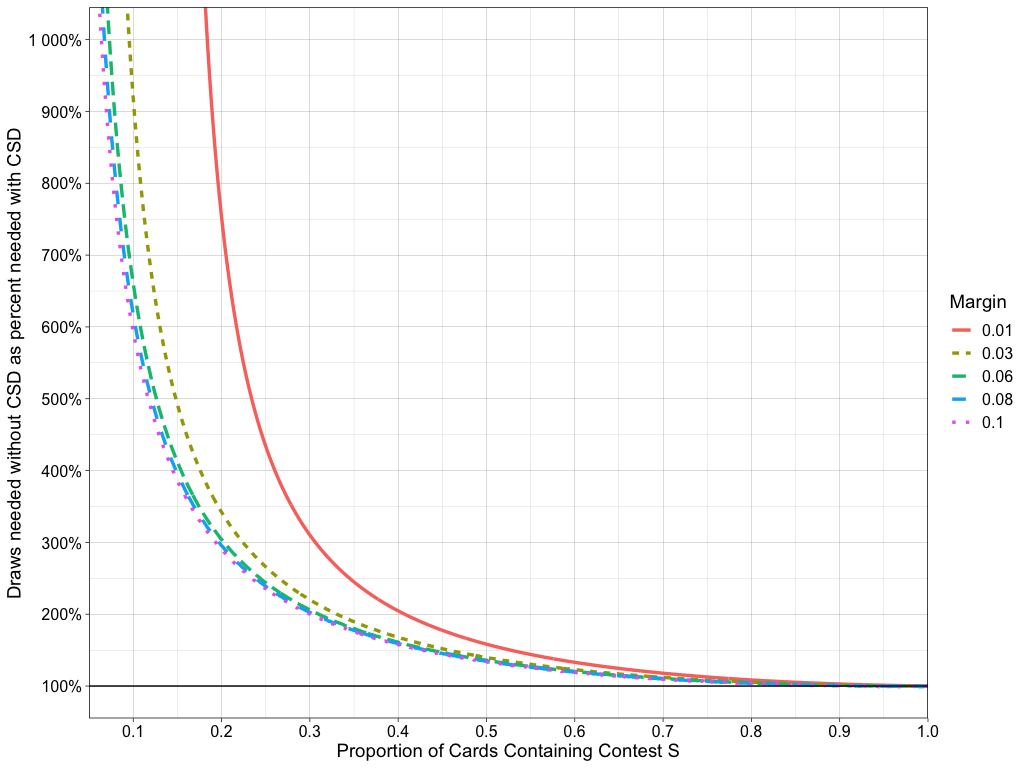}
    \caption{Percentage increase in the number of draws required for an RLA without \CSD\ compared to an RLA with \CSD. 
    Partially diluted margins with $m_B = m_S$ are given in the legend.
    Risk limit is 5\%; the audit method is S4.
    The $y$-axis shows expected draws without \CSD\ as a multiple of the draws needed with \CSD. 
    For example, for a partially diluted margin of 0.03, an audit without \CSD\ will require inspecting about 5 times as many ballots as an audit that uses \CSD\ if the small contest is on
    15\% of the ballot cards. The $y$-axis is truncated at 1,000\%. 
    The $x$-axis ranges over a grid of proportions $p$ of ballot cards containing the small contest, from 1 in 10 ($p = 0.1$) to every ballot card ($p = 1.0$).}
    \label{fig:equal_margin_pct}
\end{figure}

Figure~\ref{fig:equal_margin_pct} plots the number of cards needed without \CSD\ as a percentage of the number needed with \CSD. Without \CSD\ we need to inspect substantially more ballots when $p$ is small. 
For very small $p$, an RLA using S4 is inefficient:
a full hand count will be less work than sampling.

\begin{figure}[!ht]
    \centering
    \includegraphics[width = \textwidth]{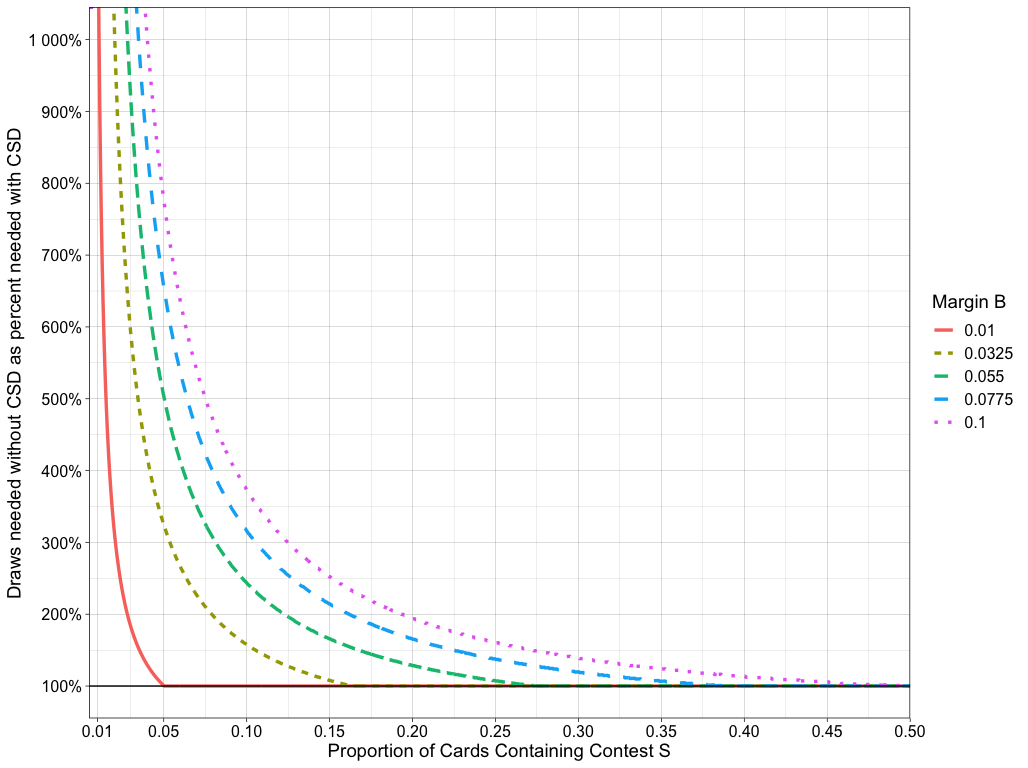}
    \caption{Draws needed to confirm the outcomes in both contest $B$ and contest $S$ when the reported outcomes are correct, as a function of the partially diluted margin in contest $B$. 
    Results are for the S4 auditing method and a risk limit of 5\%,
    on the assumption that
    the rate of 1-vote overstatement errors is $0.001$.
    The partially diluted margin for contest $S$ is fixed at $0.2$. The $y$-axis gives the expected number of draws that are needed when we do not have card-style data (\CSD) as a multiple of the number of draws needed when we do. 
    For example, for a partially diluted margin of $0.055$, when
    the smaller contest is on 5\% of the ballot cards, 5 times as many draws are needed if we do not have \CSD{}. 
    The $y$-axis is truncated at 1,000\%. 
    The $x$ axis ranges over a grid of proportions $p$ of ballot cards containing contest $S$, from 1 in 100 to 1 in 2.}
    \label{fig:unequal_margin_pct}
\end{figure}

Figure~\ref{fig:unequal_margin_pct} is similar to Figure~\ref{fig:equal_margin_pct} except $m_S$ is set to $0.1$ and $m_B$ varies from $0.01$ to $0.1$. When $p m_S > m_B$, there is no penalty for not having \CSD.

Recall that the number of cards that must be audited is 
$\rho$ divided by the fully diluted margin, so the audit of contest $B$ will examine $\frac{\rho}{m_B}$ cards and the audit of contest $S$ will (on average) examine $\frac{\rho}{m_S} - \frac{p \rho}{m_B}$ additional cards, 
for a total of $(1-p)\frac{\rho}{m_B} + \frac{\rho}{m_S}$ cards.
On the other hand, without \CSD\ we would have had to examine $\frac{\rho}{p m_S}$ ballots. 
If $m_B \geq p m_S$, \CSD\
reduces the workload by 
$$\frac{\rho}{p m_S} - \left ( (1-p)\frac{\rho}{m_B} + \frac{\rho}{m_S} \right )$$
on average.\footnote{%
The sample size for
auditing contest $B$ is fixed, as is the sample size for auditing contest $S$, but the overlap
of the two samples is random.
The expected overlap is $p \rho/m_b$.
}

\section{Multi-Card Ballots}
\label{sec:multicardballots}
Now suppose that each ballot consists of $c>1$ cards. 
For simplicity, suppose that every voter casts all $c$ cards of their ballot.
Contest $B$ is on all $N$ \emph{ballots} and on $N$ of the $Nc$ \emph{cards}.
Contest $S$ is on $Np$ of the $Nc$ cards.

Suppose there are $N = \mbox{10,000}$
ballots, $p = 0.1$, and $m_B = 0.1 = m_S$,
the risk limit $\alpha = 0.05$, and we assume that the sample will reveal 1-vote overstatement errors at a rate of $0.001$, as before.
Recall that for $c = 1$, we had to sample 721~cards without \CSD\ and 122 with \CSD.

For $c=2$, without \CSD\ the fully diluted margins are
$$
\frac{M_B}{cN} = \frac{1}{c} m_B  = 0.1/2 = 0.05
$$
for contest $B$ and 
$$
\frac{M_S}{cN} = \frac{p}{c} m_s = 0.1 \times 0.1/2 = 0.005
$$ 
for contest $S$, so the audit will examine 
$\rho/0.005 = \mbox{1,712}$ 
cards.

For $c = 5$, the fully diluted margins become 
$$\frac{1}{c} m_B = 0.1/5 = 0.02$$ 
and 
$$\frac{p}{c} m_S  = 0.1 \times 0.1/5 = 0.002
$$ 
without \CSD, so the audit will examine 9,775~cards.

If we had \CSD, we would only need to sample 122~cards as before,
no matter how large $c$ is, if every card that contains 
contest $S$ also contains $B$. 
If contests $B$ and $S$ are on different cards then with \CSD\ we would need to sample $64 + 64 = 128$ ballot cards, because no card that contains $S$ also contains $B$.

As the number $c$ of cards per ballot increases, the sample size without \CSD\ grows in proportion, but the sample size with \CSD\ stays constant: having \CSD\ saves more
work the more cards per ballot there are.

If we do not have \CSD\ we need to examine 
$$
\max \left (\frac{c \rho}{m_B}, \frac{c \rho}{p m_S} \right )
$$ 
cards. 
Suppose $m_B > p m_S$, so we need to examine $\frac{c \rho}{p m_S}$ ballot cards if we do not have \CSD.
With \CSD\ the audit of contest $B$ will examine $\frac{\rho}{m_B}$ cards and the audit of
contest $S$ will examine either an additional $\frac{\rho}{m_S}$ ballot cards if it is on a different ballot card from contest $B$ or $\frac{\rho}{m_S} - p\frac{\rho}{m_B}$ (on average) cards if every card that contains $S$ also contains $B$.
This results in the following expression for the difference in the number of draws with and without \CSD:

\begin{align*}
    \left ( \frac{c \rho}{p m_S} \right ) - \left ( \frac{\rho}{m_B} + \frac{\rho}{m_S} \right ) & ~\text{ if contests $B$ and $S$ are on different cards.}\\
    \left ( \frac{c \rho}{p m_S} \right ) - \left ( (1-p) \frac{\rho}{m_B} + \frac{\rho}{m_S} \right ) & ~\text{ if contests $B$ and $S$ are on the same cards.}
\end{align*}

\begin{figure}[!ht]
    \centering
    \includegraphics[width = \textwidth]{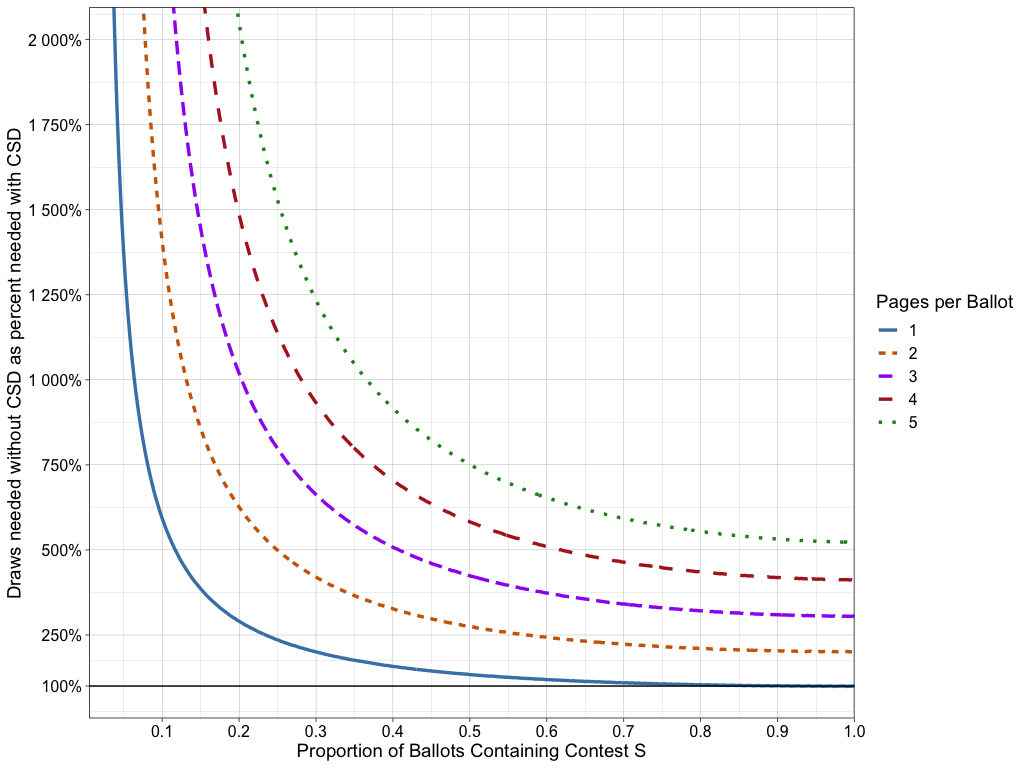}
    \caption{Draws needed to confirm the outcome in both contests ($B$ and $S$) using the S4 method at risk limit $\alpha = 0.05$ without \CSD\ as a multiple of the number of draws needed with \CSD\ ($y$-axis), truncated at 2,000\%. 
    The contests appear on the \textit{same} ballot-card and both partially diluted margins are fixed at 0.1. The lines indicate multiple needed; different lines correspond to different numbers of cards per ballot. 
    The $x$-axis ranges over the proportion of ballots (not ballot cards) containing contest $S$, from 1 in 100 ($p=0.01$) to all ballots ($p = 1.0$).}
    \label{fig:multiple_same_pages}
\end{figure}

Figure~\ref{fig:multiple_same_pages} plots the sample size needed without \CSD\ as the percentage needed with \CSD\ in an election with a multi-card ballot. The number of cards per ballot, $c$, ranges from 1 to 5, while the partially diluted margins are fixed at $m_b = m_S$. Contests $B$ and $S$ appear on the same card.

\begin{figure}[!ht]
    \centering
    \includegraphics[width = \textwidth]{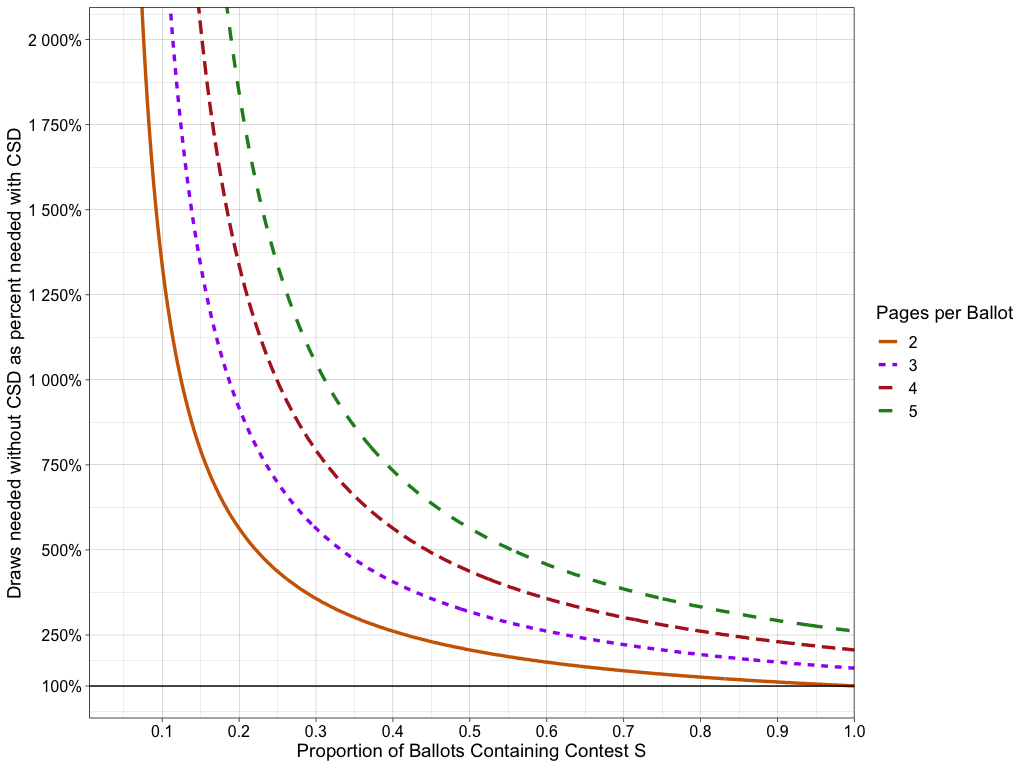}
    \caption{Draws needed to confirm outcome in both contests ($B$ and $S$) when we do not have \CSD\ expressed as a percentage of draws needed with \CSD\ ($y$-axis), truncated at 2,000\%. The contests appear on \textit{different} pages, while both partially diluted margins are fixed at 0.1. The lines indicate percentage of draws needed, colored by the number of pages in the ballot. The $x$-axis ranges over the proportion of ballots (not ballot cards) containing contest $S$, from 1 in 100 ($p=0.01$) to all ballots ($p = 1.0$).}
    \label{fig:multiple_different_pages}
\end{figure}

Figure~\ref{fig:multiple_different_pages} plots the sample size needed under the same set-up as Figure~\ref{fig:multiple_same_pages}, except that $B$ and $S$ appear on different cards. 

\section{Ballot-polling audits} \label{sec:bpa}
We have focused on comparison audits because we expect \CSD\ will be derived from CVRs;
when the voting system can produce CVRs linked to
physical ballot cards, ballot-level comparison audits are possible and save a substantial amount of work compared to ballot-polling audits.

Jurisdictions that cannot conduct ballot-level comparison audits can still conduct ballot-polling audits. 
We show here that \CSD\ (derived, for instance, by physically sorting ballot cards) can yield similar savings for ballot-polling audits.
Whether the savings in audit effort is worth the effort of sorting the cards will depend on the jurisdiction's logistics, details of the contests under audit (including the partially diluted margins), and the number of cards per ballot.

For ballot-polling audits, the expected sample size scales like the square of the reciprocal of the margin and
linearly in the reciprocal of the fraction of ballot cards that contain the contest. 
In other words, without \CSD\ the sample sizes expected for ballot-polling audits scale linearly in $1/p$, the proportion of cards containing the smaller contest, and
linearly in $c$, the number of cards per ballot---just
as comparison audits do.

Consider again auditing
contests $B$ and $S$ with margins $M_B$ and $M_S$ (in votes), 
$N$ ballots cast each consisting of $c$ cards, 
contest $B$ on $N$ of the $Nc$
cards and $S$ is on $pN$ of the cards. 
For simplicity, assume that the contests are two-candidate 
plurality contests with no invalid votes; with small changes to the notation, the result can be generalized.

For the BRAVO method for ballot polling \cite{lindemanEtal12}, the expected sample size is approximately $\frac{2 \ln(1/\alpha)}{m^2}$ where $m$ is the margin as a fraction (the margin in votes,
divided by the number of votes cast for the winner or the loser).

Suppose we know which \emph{ballots} contain $S$
but not which particular \emph{cards} contain $S$,
and that the $c$ cards comprising each ballot are kept
in the same container.
This is an idealization of precinct-based voting where
each voter in a precinct gets the same ballot style
and casts all $c$ cards of the ballot.
If voters in that precinct are eligible to vote in $S$, a
fraction $1/c$ of the cards in the container will have contest $S$; otherwise, none of the cards in the container will have $S$.

This lets us target the sampling for auditing contest $S$, but only partially: we can reduce the sampling universe for contest $S$ from the original population of $Nc$ cards to a smaller population of $pNc$ cards, of which $pN$ actually contain contest $S$.
The sample for $B$ would be drawn from all $Nc$ cards, of which $N$ contain contest $B$.
The difference in expected sample sizes compared to ``blind'' ballot polling with
no \CSD\ is
\begin{align*}
    2c \ln (1/\alpha) \left ( \frac{1}{p m_S^2} -  \frac{1}{m_B^2} - \frac{1}{m_S^2} \right ) & ~\text{ if contests $B$ and $S$ are on different cards.}\\
    2 c \ln(1/\alpha) \left ( \frac{1}{p m_S^2} -  (1-p) \frac{1}{m_B^2} - \frac{1}{m_S^2} \right )& ~\text{ if contests $B$ and $S$ are on the same card.}
\end{align*}

For a risk limit of 5\% and a margin of 10\%, using BRAVO, we would expect to sample 608~cards if we could target the sample. 
Consider the example $N = \mbox{10,000}$, $p = 0.3$, $M_B = 0.1 = M_S$. 
If $c = 2$, absent \CSD{} we would expect to sample $(2/0.3)\times 608 = \mbox{4,053}$ cards.
With partial \CSD{}, but no information about which contests are contained on an individual card, we would expect to sample $(1 - 0.3)2\times 608 + 2\times 608 = \mbox{2,067}$ cards if the contests are on the same card and $2\times 608 + 2\times 608 = \mbox{2,432}$ ballot cards if the contests are on different cards.
In either case, using \CSD\ reduces the sample size by roughly half.

Thus, information about which containers have which
card styles---even without information about which cards contain which contests---can still yield substantial efficiency gains for ballot-polling audits.

\section{Case studies}
\label{sec:case_studies}

We will give numerical examples for Inyo County, California (a small county), and Orange County, California (a large county), 
both of which have conducted several RLA pilots or binding RLAs.

\subsection{Inyo County, California}

Inyo County (population 18,546 according to the 2010 U.S.~Census) is in eastern California. 
The county conducted a pilot RLA of an April 2018 special election using the S4 method \cite{foote18a}
(it has conducted other RLAs as well, including RLAs of 7~contests in the 2020
general election).

In the June 2018 election each ballot consisted of two cards ($c = 2$).
We consider two contests on this ballot: 
Supervisor District~1 (contest $S$) and U.S.~Senator (contest $B$).
(We shall
pretend that the Senate contest was entirely contained in Inyo County.)
Of the 5,919 ballots cast in the June 2018 election, 1,435 ($p = 0.24$) contained the Supervisor District~1 contest. 
This contest had a partially diluted margin of $m_S = 36/\mbox{1,435} = 0.025$.
In California primaries, the top \textit{two} candidates with the most votes (from either party) advance to the general election, so the margin between the 2nd place candidate and the 3rd place candidate drives the audit effort. 
All voters in Inyo County were eligible to vote in the U.S.~Senate contest. 
Diane Feinstein received the most votes, 1,555, followed by James Bradley with 639, and Paul Taylor with 517. 
We consider confirming that Feinstein and Bradley received more votes than Taylor.
The partially diluted
margin for this contest is 
$$m_B = (639 - 517)/\mbox{5,919} = 0.02 > 0.006 = pm_S.$$

First, consider a comparison audit.
Without \CSD, we would expect to sample 3,734 cards.
The Supervisor District~1 and U.S.~Senate contests were not on the same card, so with \CSD\ we would expect to sample 588~cards, smaller by about $\mbox{3,734} - 588 = \mbox{3,146}$ cards than
an audit without \CSD{}.

Suppose instead we audit using BRAVO, a ballot-polling method.
Without \CSD\ we would expect to audit $\mbox{2,796} \times 2$ ballot cards to verify contest $B$, but to confirm the outcome of contest $S$ would essentially require a full hand count
of the votes on all $\mbox{11,838}$ 
cards. 
If we knew which containers had cards that include contest
$S$, we would expect to audit $\mbox{2,796} \times 2 + \mbox{1,435} \times 2 = \mbox{8,462}$ cards.
(While this is a substantial reduction, it is probably
more efficient to conduct a full hand count than to examine a majority of ballot cards selected randomly.)

\subsection{Orange County, California}
Orange County, in Southern California, is much larger (3.017 million residents according to the 2010 U.S.\ Census).
The ballots in their 2018 election consisted of $c=2$ cards. 
A BRAVO RLA of three of the five countywide contests in 2018 was
conducted \cite{singermcburnett18a}.

Suppose we wanted to audit the Senate (Feinstein v De Leon) and 45th District Congressional (Porter v Walters) contests in Orange County from the November 2018 election.
(As before, we shall pretend for the purpose of illustration
that the senatorial contest is entirely contained in Orange County.)
All voters were eligible to vote in the Senate contest. 
Of the 1,106,729 ballots cast, 312,700 (28.25\%) included the 45th District Congressional contest.
In this case $p \approx 0.2825$, $m_S = 0.04$, $m_B = 0.073$, and $c = 2$, so
$$m_B = 0.073 > 0.012 = p m_S.$$

First consider a ballot-level comparison audit.
Without \CSD, we would expect to have to sample 1,452 cards.
With \CSD, we would expect to have to sample 249 cards if the contests were on different cards, and 225 if they were on the same card.
These two contests were in fact on the same ballot card, so using \CSD\ would decrease the expected number of
audited cards by $\mbox{1,452} - 225 = \mbox{1,227}$.

Suppose instead we audit using the BRAVO
ballot-polling method.
If we did not have \CSD, we would expect to have to sample $2*\mbox{3,671}/0.2825 = \mbox{25,989}$ ballot cards.
If ballots were organized by ballot style (but not card style), we would expect to sample 
$(1 - 0.2825) \times 2 \times 948 + 2 \times \mbox{3,671} = \mbox{8,702}$ cards.

Table~\ref{tab:examples} summarizes the results of this section, the expected percentage reduction in number of ballot cards drawn with \CSD\ versus without \CSD.

\begin{table}[!ht]
\centering
\begin{tabular}{|l|l|l|}
\hline
 & Comparison Audit & Ballot-Polling Audit \\ \hline
Inyo County         & 84\%                & 29\%                    \\ \hline
Orange County       & 85\%                & 67\%                    \\ \hline
\end{tabular}
\\
\caption{Expected percentage reduction in the sample size required with \CSD\ versus without \CSD\ to audit the contests described in Section~\ref{sec:case_studies}.}
\label{tab:examples}
\end{table}

\section{Implementation}
\label{sec:implementation}

\begin{figure}[!ht]
    \centering
    \includegraphics[width = \textwidth]{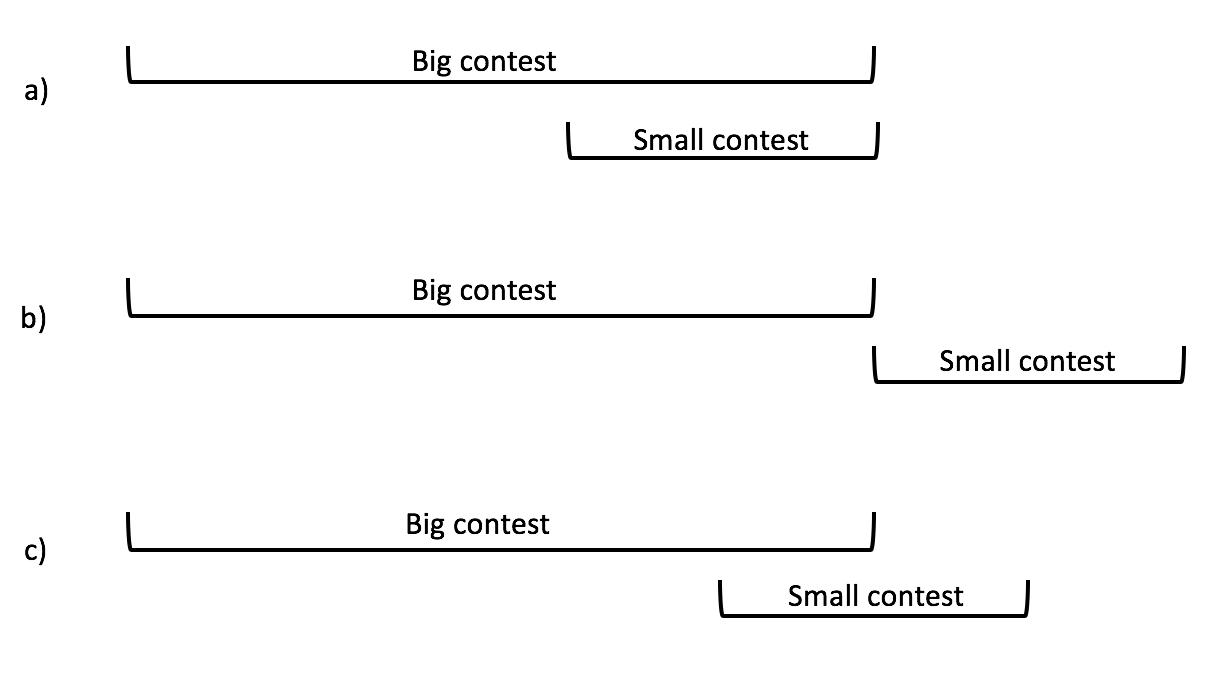}
    \caption{Overlap of card styles in 
    3~hypothetical elections with 
    2~contests (one big and one small). Case $a$ represents complete nesting, where uniform draws from the big contest yield uniform draws from the small contest. If all ballots consist of a single ballot card and the big contest is on all ballots, the situation is case $a$. Case $b$ represents a scenario where ballot cards that contain the big contest do not contain the small contest. 
    Auditing the big contest tells us nothing about the small contest. In case $c$, the small contest sometimes appears on the card for the big contest and sometimes not. Drawing ballot cards uniformly from the big contest does not sample from the entire small contest.}
    \label{fig:ballot_bins}
\end{figure}

To minimize the number of cards a
risk-limiting audit of multiple
contests needs to inspect,
we would like to be able to use any card in the audit sample
to audit every contest that card contains.
Standard methods for ballot-level comparison audits or ballot-polling audits then require that the intersection of the sample with the cards that contain each contest is a uniform random sample from that contest (either with or without replacement, as appropriate for
the auditing method).
When the contests are on some of the same cards,
this requires particular care.
Figure~\ref{fig:ballot_bins} sketches the possibilities.

\begin{figure}[!ht]
    \centering
    \includegraphics[width = \textwidth]{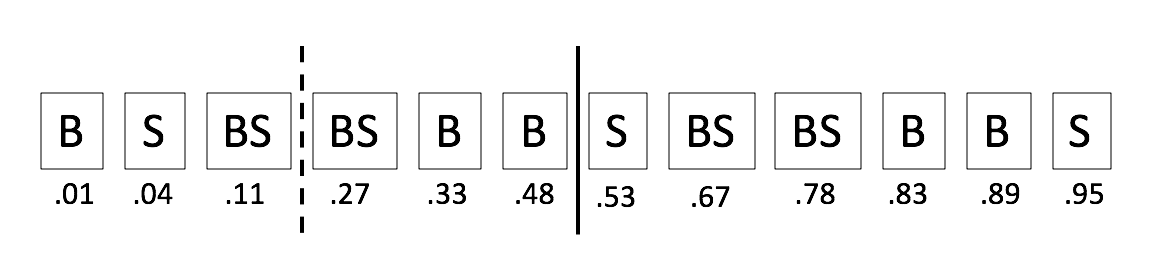}
    \caption{Illustration of consistent random sampling without replacement to audit 2 contests, $B$
    and $S$.
    Contest $B$ appears on 9~cards and $S$ appears on 7~cards; 12~cards were cast in all.
    There is partial overlap (case~c of Figure~\ref{fig:ballot_bins}): contest $B$ appears alone on 5~cards (labelled ``B''), the contests appear together on 4~cards (labelled ``BS''), and contest $S$ appears alone on 3~cards (labelled ``S''). 
    To perform consistent sampling, each card is assigned a random number in $[0,1]$ independently across cards.
    The number appears below each card;
    without loss of generality, we have sorted the cards by increasing order of the random number. 
    Suppose the audit requires inspecting 5~cards containing $B$ and 2~cards containing $S$.
    The 5~sampled cards containing $B$ are the cards containing $B$ that were assigned the 5~smallest numbers, i.e., all cards containing $B$ to the left of the second (solid) vertical bar.
    The 2~sampled cards containing $S$ are the cards containing $S$ that were assigned the 2~smallest random numbers, i.e., all cards containing $S$ to the left of the first (dashed) vertical bar.
    This approach ensures that if we look at a card to audit one contest and the other contest is also on the card, it can be used in the audit of that contest, too:
    the sample is uniformly distributed on subsets of the population.
    This approach can be generalized to sampling with replacement; see \cite{rivest18b}.}
    \label{fig:consistent_sampling}
\end{figure}

One method for ensuring uniformity on overlapping
subsets is called \emph{consistent sampling}.
Consistent sampling can be done with replacement \cite{rivest18b} 
or without replacement \cite{broder97a} 
\cite{broder97b}.
It is simpler without replacement, which also leads to more
efficient RLAs \cite{stark20a}.
Figure~\ref{fig:consistent_sampling} illustrates 
consistent sampling with a toy example of auditing two contests in a small election. 

Here we show how to audit $K$ contests of different sizes using \CSD\ and consistent sampling. 
We assume that (1)~there has been a \emph{compliance audit}
to establish that the paper trail is trustworthy
\cite{benalohEtal11,starkWagner12,starkEBE18,appelStark20}; (2)~there is an upper bound on the number of cards that contain each contest under audit, obtained, for example, from pollbook signatures
and other administrative records; and 
(3)~the underlying auditing method  
(comparison vs ballot-polling, and the ``risk function''),
sampling method, and risk limits ($\{\alpha_1,.. \alpha_K\}$) have been chosen, along with
rules for picking the initial sample size(s)
$\{S_1,...,S_K\}$ 
and for increasing the sample size
if the audit does not confirm the outcome
with the initial sample.

\begin{enumerate}
    \item If there are more CVRs that contain any contest than the upper bound on the number of cards that contain the contest, stop: the contest outcome cannot be confirmed.
    \item If the upper bound on the number of cards that contain a contest is greater than the number of CVRs
    that contain the contest, create a corresponding set of ``phantom'' CVRs as described in section~3.4 of
    \cite{stark20a}.
    \item If the upper bound on the number of cards that contain a contest is greater than the number of physical cards whose locations are known, create enough ``phantom'' cards to make up the difference.
    \item Assign a $U[0, 1]$ pseudo-random number to every 
    ballot card that contains one or more contests under audit (including ``phantom'' cards), using a high-quality PRNG \cite{OttoboniStark19}. 
    For sampling without replacement, assign one number to each card; for sampling with replacement, assign 
    several, as described in \cite{rivest18b}.
    
    \item Initialize $\mathcal{A}$ to be the set of contests under audit: $\mathcal{A} \leftarrow \{1, \ldots, K\}$.
    
    \item While $\mathcal{A}$ is not empty:
    
    \begin{enumerate}
        \item Pick the sample sizes $\{S_k\}$ for $k \in \mathcal{A}$ for this round of sampling. 
        \item Choose thresholds $\{t_k\}_{k \in \mathcal{A}}$ so that $S_k$ ballot cards containing contest $k$ have numbers less than or equal to $t_k$.
        \item Retrieve any of the corresponding ballot cards that have not yet been audited and inspect them manually.
        If there is no CVR for the ballot, treat the CVR as if it recorded a non-vote in every contest still under audit.
        If a ballot card cannot be found or if it is a phantom card, treat it in the way that casts the most doubt on the outcome of every audited contest it was supposed to contain (see Section~3.4 of \cite{stark20a}).\footnote{Some cards may be selected for auditing more than one contest. If the sample is drawn with replacement, the same card may be selected more than once. Such cards are only manually inspected once, even though their data might be re-used.}
        
        \item  Use the data from the previous step to update the 
        measured risk for every contest $k \in \mathcal{A}$.
    
        \item Remove from $\mathcal{A}$ all $k$ that have met their risk limits. 
    \end{enumerate}
\end{enumerate}

\section{Conclusions}
\label{sec:conclusions}

Card-style data (\CSD) can dramatically increase the efficiency of risk-limiting audits. 
For the super-simple simultaneous single-ballot audit of \cite{stark10d} and for ballot-polling audits using BRAVO \cite{lindemanEtal12},
the expected reduction in sample size can easily be
several orders of magnitude, depending
on the range of sizes and margins of the contests
under audit and the number of cards per ballot.

When a contest is on only a small fraction of the cards cast in an election and the sample is drawn from all cast cards, auditing the contest can be nearly as much work as a full manual tally involving \emph{all} cast cards, not just those
that contain the contest---even if the margin in the contest is large.
In contrast, if there are \CSD\ so that the sample can be drawn from just those cards that (reportedly) contain the contest, auditing small
contests can be quite efficient. 

Jurisdictions that perform RLAs should consider using \CSD\ to reduce the workload of auditing small contests and contests on multi-card ballots: the savings can be substantial.
Jurisdictions that can conduct ballot-level comparison audits (i.e., jurisdictions with voting systems that can export CVRs linked to physical ballots) can construct \CSD\ with no
additional effort, because the CVRs
contain \CSD---albeit, possibly erroneous.
However, to ensure that errors in the CVRs resulting in errors in \CSD\
do not compromise the risk limit, jurisdictions also need an upper bound on the
total number of ballot cards that contain each contest,
information that can be derived from pollbooks and related administrative 
voter records.

\section*{Acknowledgements}
Thank you to Kammi Foote and Neal Kelley for providing data for the Inyo and Orange County case studies.
Thanks also to the anonymous reviewers for their helpful feedback.

Jacob Spertus' work was supported by a National Science Foundation (NSF) Graduate Research Fellowship (DGE 1752814). Amanda Glazer's work was supported by an NSF grant (DMS RTG \#1745640).


\bibliographystyle{splncs04}
\bibliography{main}

\end{document}